\documentclass[]{IEEEtran}
\usepackage{graphicx}
\usepackage{float}
\usepackage{dblfloatfix}
\usepackage{xcolor}
\usepackage{placeins}
\usepackage{scrextend}
\usepackage[margin=0.625in]{geometry}
\usepackage{anyfontsize}
\usepackage{titlesec}
\usepackage[font=small]{caption}
\makeatletter

\def\footnoterule{\relax%
  \kern-5pt
  \hbox to \columnwidth{\hfill\vrule width 0.8\columnwidth height 0.4pt\hfill}
  \kern4.6pt}
\makeatother

\usepackage{multirow}

\definecolor{titleBlue}{RGB}{0,0,153}
\definecolor{titleRed}{RGB}{204,0,0}
\definecolor{abstrRed}{RGB}{192,0,0}

\IEEEoverridecommandlockouts

\IEEEaftertitletext{\vspace{-2\baselineskip}}
\begin{document}



\title{Logical and Physical Reversibility of Conservative Skyrmion Logic} 

\author{
\IEEEauthorblockN{
    Xuan Hu\IEEEauthorrefmark{1},
    Benjamin W. Walker\IEEEauthorrefmark{1},
    Felipe Garc\'ia-S\'anchez\IEEEauthorrefmark{2},
    Alexander J. Edwards\IEEEauthorrefmark{1},
    Peng Zhou\IEEEauthorrefmark{1},\\
    Jean Anne C. Incorvia\IEEEauthorrefmark{3},
    Alexandru Paler\IEEEauthorrefmark{4},
    Michael P. Frank\IEEEauthorrefmark{5},
    Joseph S. Friedman\IEEEauthorrefmark{1}
}
\vspace{2 pt}

\IEEEauthorblockA{
    \IEEEauthorrefmark{1}University of Texas at Dallas, Richardson, TX\\
    \IEEEauthorrefmark{2}Universidad de Salamanca, Salamanca, Spain\\
    \IEEEauthorrefmark{4}Aalto University, Espoo, Finland\\
    \IEEEauthorrefmark{3}University of Texas at Austin, Austin, TX\\
    \IEEEauthorrefmark{5}Sandia National Laboratories, Albuquerque, NM
}

}

\maketitle
\begin{abstract}
Magnetic skyrmions are nanoscale whirls of magnetism that can be propagated with electrical currents. The repulsion between skyrmions inspires their use for reversible computing based on the elastic billiard ball collisions proposed for conservative logic in 1982. Here we evaluate the logical and physical reversibility of this skyrmion logic paradigm, as well as the limitations that must be addressed before dissipation-free computation can be realized.
\end{abstract}
\renewcommand\IEEEkeywordsname{Keywords}



  
  


\section{Introduction}

The conventional {\em non-reversible} paradigm for general digital computing will eventually approach fundamental thermodynamic limits on its energy efficiency, which stem ultimately from the fact that this standard paradigm relies primarily on operations that systematically discard correlated logical information, and therefore increase entropy (\cite{Fra05,Fra18,FS21}). For example, a typical digital logic architecture in CMOS destructively overwrites the output of each active logic gate with a new bit value on each clock cycle. It has been known ever since the field of the thermodynamics of computation was first elucidated by Landauer and Bennett (\cite{Lan61,Ben73,Ben82,Ben03}) that general digital computing
technologies can only avoid the consequent limits on their energy efficiency if they are instead re-architected on the basis of {\em logically reversible} operations that are implemented (at the device and circuit level) in a nearly {\em thermodynamically reversible} way \cite{Fra17}.  

Logical reversibility requires logical operations to conserve information content, and is a prerequisite for thermodynamic reversibility. In a thermodynamically or physically reversible system, the entropy of the environment does not increase during computation, enabling incredible energy savings.  While complete physical reversibility is not attainable in any non-equilibrium system at nonzero temperature, systems can be designed to approach these thermodynamic bounds.

Although this alternative paradigm of {\em reversible computing} can be implemented in CMOS technology (\cite{YK93,Fra+20,FBTMA20}), there is also a need to 
explore a range of other, novel device technologies to determine whether they may potentially exhibit advantages for reversible computing compared to CMOS \cite{Fra99}.
In the line of work reported here, magnetic skyrmions are being explored as a candidate device technology for the implementation of reversible computing.

\section{Logically Reversible Skyrmion Logic}
\begin{figure}[tb]
    \centering
    \includegraphics[width=1\columnwidth]{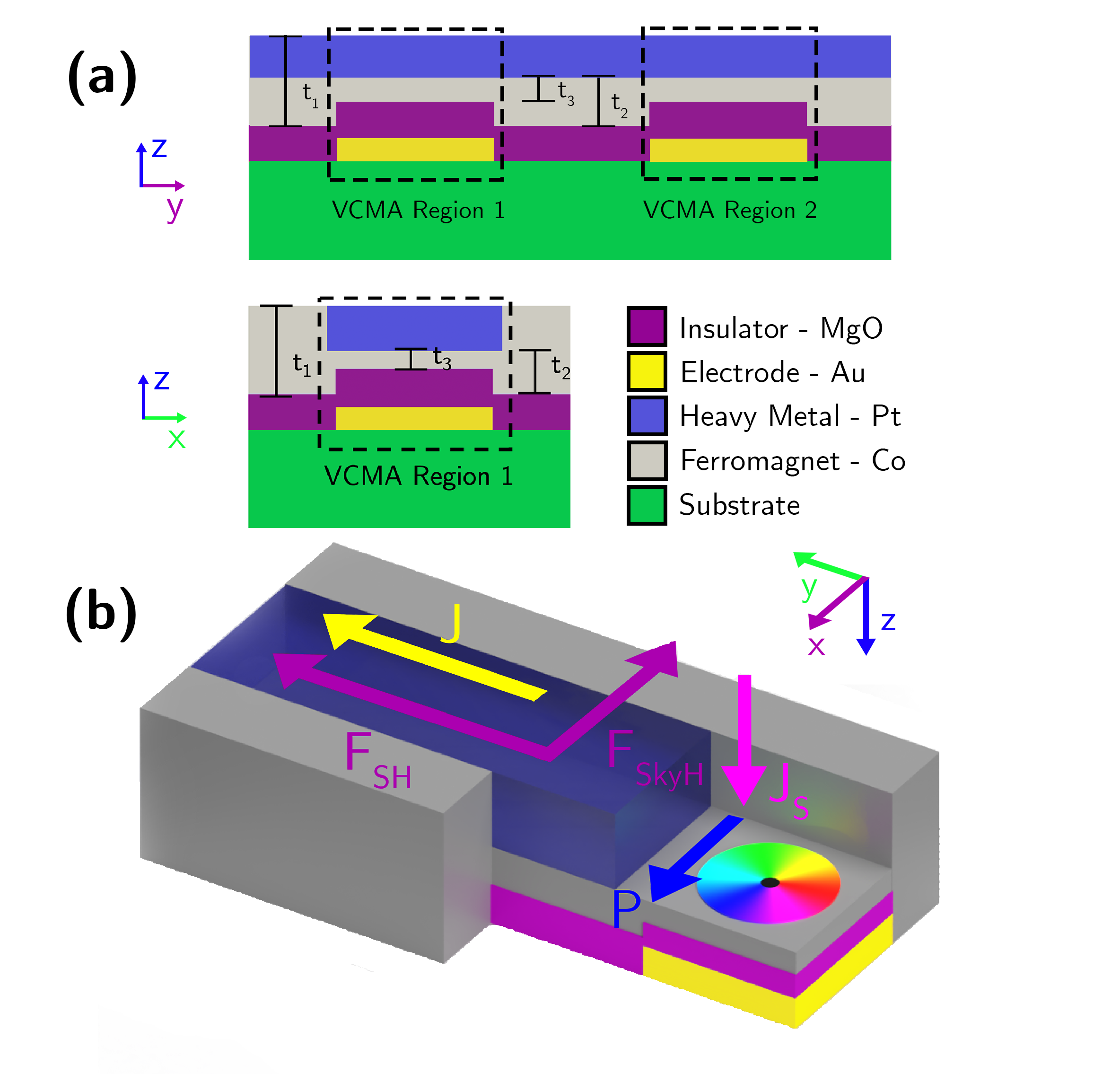}
    \caption{Skyrmion logic track structure. (a) Synchronizer cross sections in the yz and xz planes. The combined track thickness is 1.6 nm, with a track width of 20 nm. (b) A N\'eel skyrmion (colored circle) exists in the ferromagnetic layer. Injected electric current in the +y direction ($J$) induces a spin current ($J_S$)  with polarization P in the +z direction via the spin-Hall effect. This spin current produces a +y-directed spin-Hall force ($F_{SH}$), propelling the skyrmion in the +y direction. The skyrmion-Hall $(F_{SkyH})$ effect creates an x-directed force, which deviates the skyrmion from its current trajectory unless a track is present to provide repulsion. \vspace{-10pt}}
    \label{fig:structure}
\end{figure}
\begin{figure}[]
    \centering
    \includegraphics[width=\columnwidth]{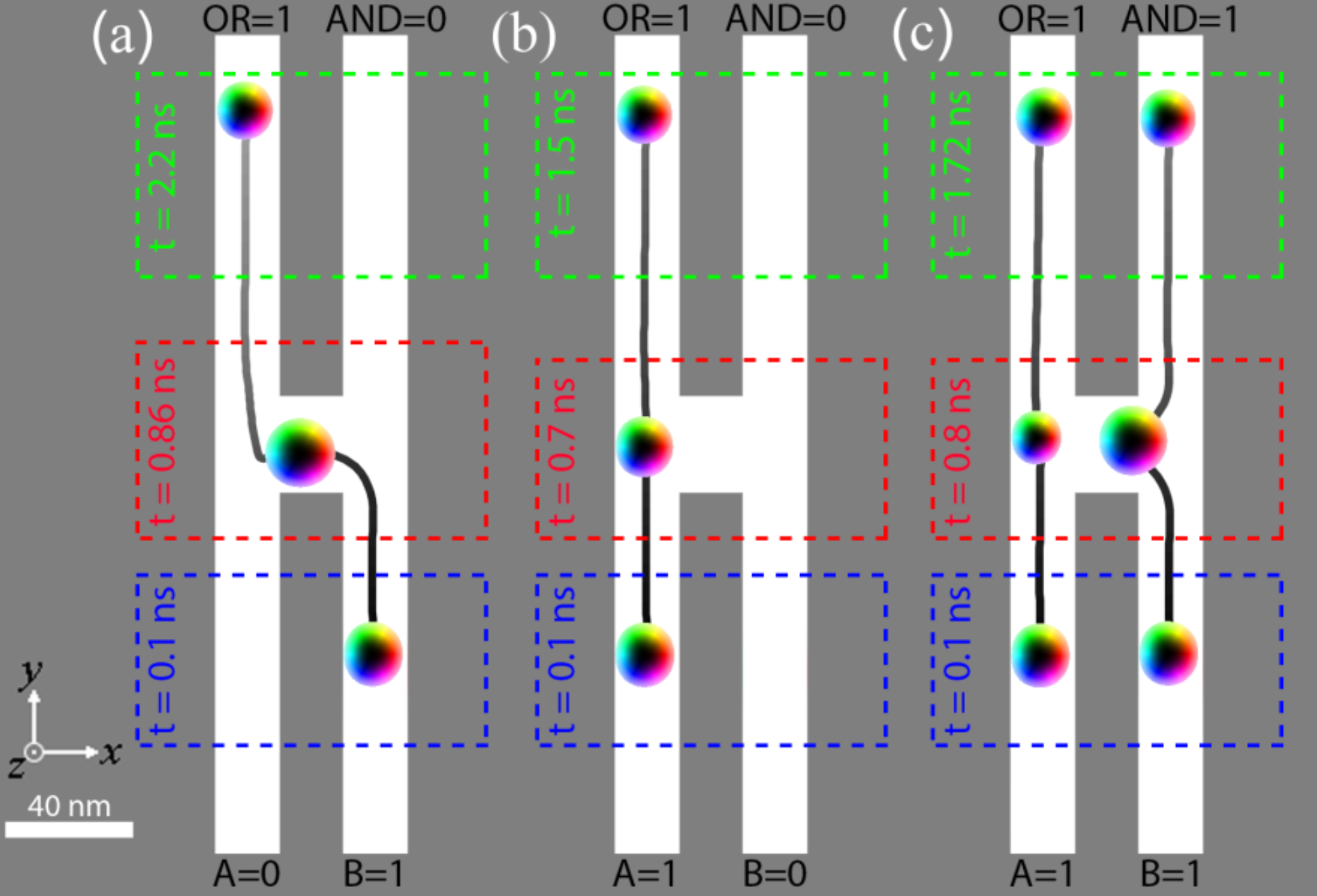}
    \caption{Micromagnetic simulation results for AND/OR gate with current ($5\times10^{10}A/m^2$) in $+y$ direction, with skyrmion trajectory represented by the black path (a) A=0, B=1 (b) A=1, B=0 (c) A=1, B=1. All simulations are performed with Mumax3 \cite{Mumax}}
    \label{fig:andor}
\end{figure}
\begin{figure}[]
    \centering
    \includegraphics[width=\columnwidth]{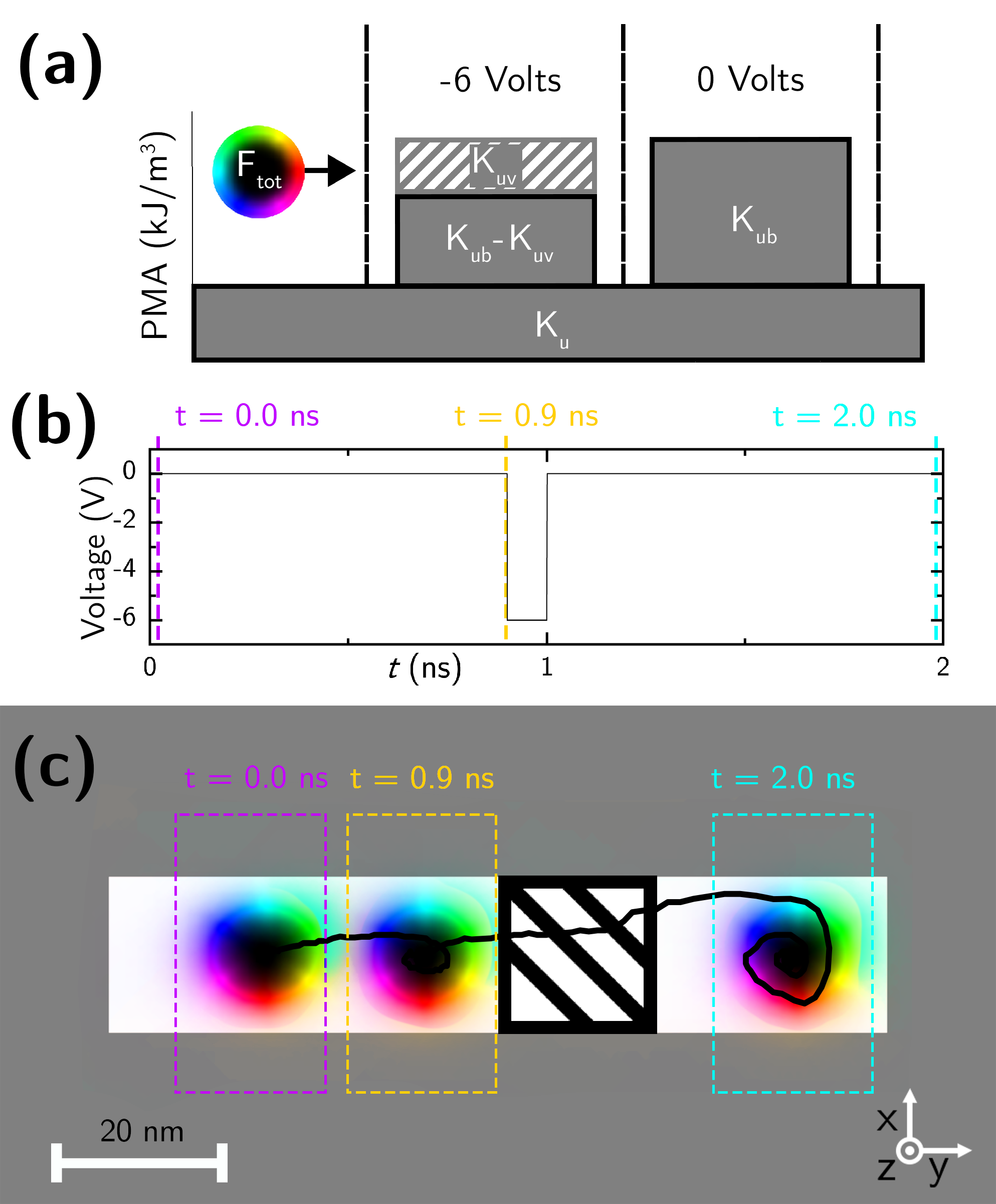}
    \caption{VCMA synchronizer with a constant injected current of $5\times10^{10}A/m^2$ in the $+y$ direction. (a) Illustration of the skyrmion synchronization method. The skyrmion is depinned when $F_{tot} >K_{ub} + K_{uv}$ (b) Voltage applied to the VCMA region over one clock cycle.  (c) Micromagnetic simulation results. Skyrmion trajectory is represented by the black line and the VCMA region is represented by the dashed square. The positive anisotropy barrier pins the skyrmion at t = 0.9 ns and the negative voltage at t = 1.0 ns reduces the barrier, depinning the skyrmion.}
    \label{fig:vcma}
\end{figure}
\begin{figure}[b!]
    \centering
    \includegraphics[width=\columnwidth]{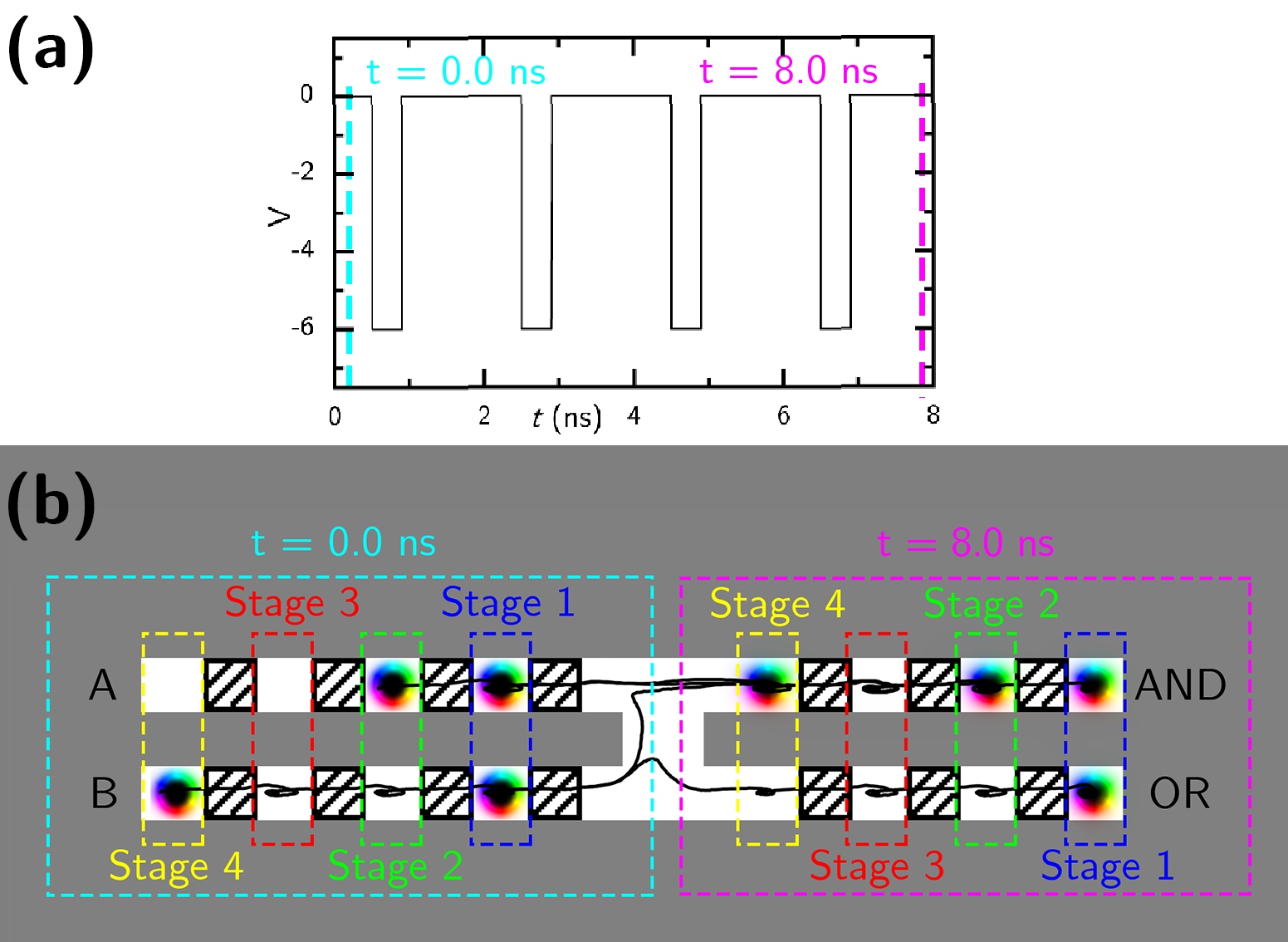}
    \caption{VCMA synchronization of pipelined skyrmion AND/OR gate, with a constant current ($5\times10^{10}A/m^2$) injected in the $+y$ direction. (a) Clock waveform applied to the VCMA synchronizers (b) Micromagnetic simulation results for pipelined AND/OR gate, demonstrating proper logical operation.}
    \label{fig:candor}
\end{figure}
\begin{figure*}[]
    \centering
    \includegraphics[width=1\textwidth]{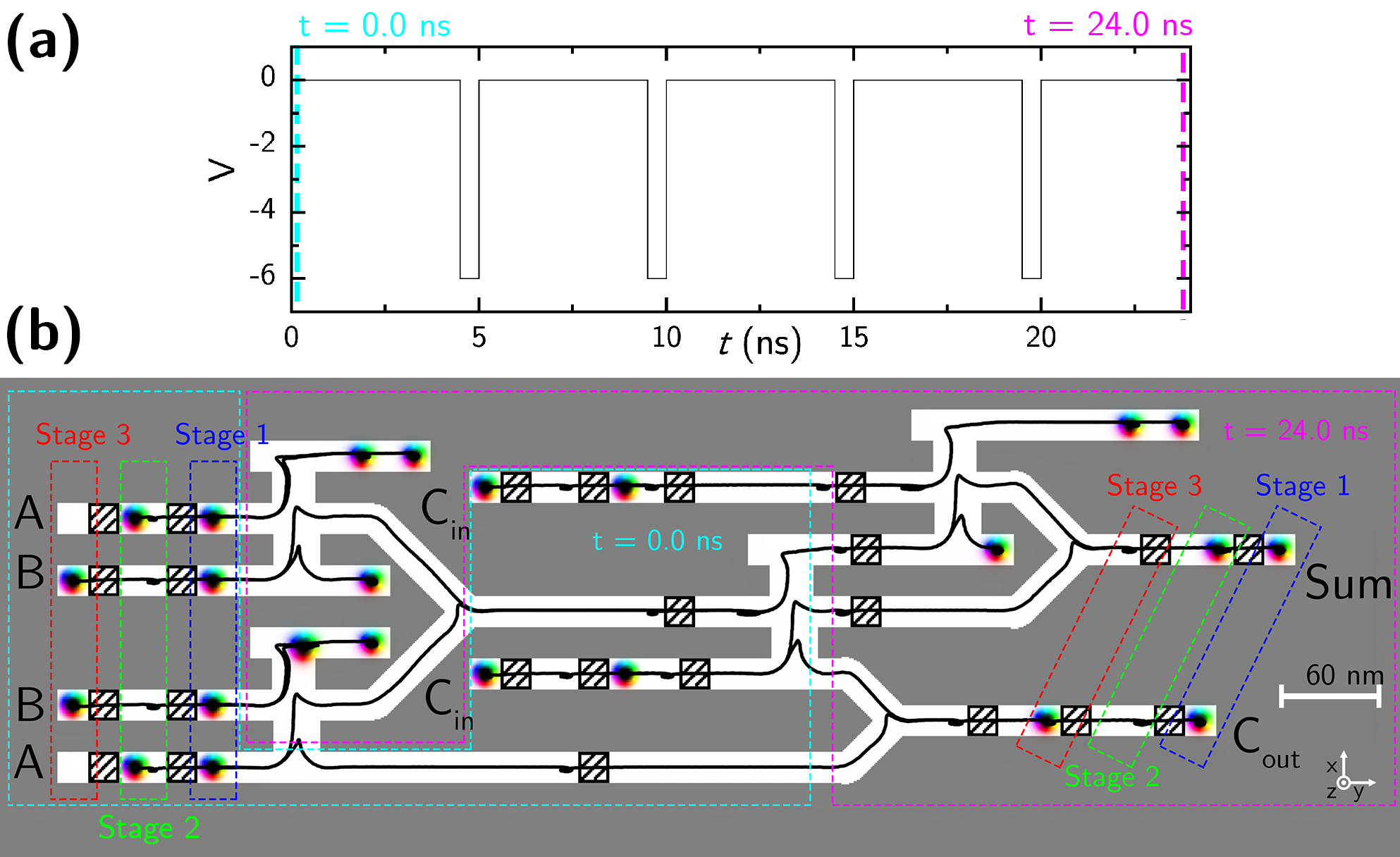}
    \caption{VCMA synchronization of pipelined skyrmion one-bit full-adder with a constant current ($5\times10^{10}A/m^2$) injected in the $+y$ direction. (a) Clock waveform applied to the VCMA synchronizers (b) Micromagnetic simulation results for pipelined one-bit full-adder, demonstrating proper logical operation.  Note that in a real system, the dead-ends would be replaced with out-of-plane tracks for cascaded logic.\vspace{0pt}}
    \label{fig:cadder}
\end{figure*}


Skyrmions are topologically stable magnetic quasiparticles that can be propelled with applied electrical current or pinned with voltage-controlled magnetic anisotropy (VCMA). Their mobility paired with the skyrmion-Hall effect and skyrmion-skyrmion repulsion allows for the construction of a skyrmion-based logic system \cite{Chauwin} based on billiard ball computing \cite{Fredkin}. Fig. \ref{fig:structure} shows the structure of such a system, where a skyrmion exists inside the Cobalt layer due to the Dzyaloshinskii–Moriya interaction between the ferromagnet and heavy metal. Encoding a logical ``1" or ``0" as the presence or absence of a skyrmion, respectively, the skyrmion logic scheme manipulates these forces to create logic functions such as the AND/OR gate, the Ressler-Feynman switch gate \cite{Ressler}, and one-bit full adders. This proposed logic scheme conserves skyrmions, and its structures can exhibit logical reversibility.

Four forces underlie this conservative skyrmion computing scheme (Fig. \ref{fig:structure}). Electronic current in the $+y$ direction moves the skyrmion in the same direction via the spin-Hall effect. The skyrmion's velocity causes it to deviate in a direction perpendicular to its motion ($-x$) through the skyrmion-Hall effect. Track-edge repulsion allows the skyrmion to move in a straight line despite the skyrmion-Hall effect, and skyrmion-skyrmion repulsion provides the billiard ball-like interaction necessary for logic \cite{Chauwin,Walker}.


The skyrmion AND/OR gate uses these forces for conditionally reversible computation (Fig. \ref{fig:andor}). Electronic current moves the skyrmions in a straight line in the $+y$ direction due to track repulsion. However, when a skyrmion in the right track reaches the gate junction, the loss of track repulsion allows it to move in the $-x$ direction into the left lane (Fig. \ref{fig:andor}a). Skyrmions in the left lane remain in the left lane, due to constant track repulsion (Fig. \ref{fig:andor}b). Importantly, when two skyrmions are synchronized in both tracks (Fig. \ref{fig:andor}c), the skyrmion-skyrmion repulsion allows for the right skyrmion to remain in the right track. The output of left track therefore represents logical OR, while the right represents logical AND.  Due to the input ambiguity of the $(OR=1, AND=0)$ case, the AND/OR gate is conditionally logically reversible \cite{Fra17} if one of the two input cases, $(A=0, B=1)$ or $(A=1, B=0)$, is excluded from operation, allowing a version of this gate to have an implementation which approaches the thermodynamic limit. However, a reversible version of this particular gate would not be useful; later we discuss a more useful reversible gate.

As skyrmions need to be synchronized when entering a gate for proper logical operation, varying skyrmion path-lengths and thermal fluctuations can result in logical errors. Through VCMA clocking \cite{Walker}, the application of an electric field (+z) via an electrode changes the perpendicular magnetic anisotropy (PMA) of the track. As skyrmions can be pinned by a PMA gradient, the voltage modulation of the electrode at the VCMA region (Fig. \ref{fig:vcma}) can be used to pin and depin skyrmions as a clocking mechanism. These synchronizers are used to create a pipelined AND/OR gate in Fig. \ref{fig:candor}.

When multiple skyrmion gates are cascaded together, complex functions can be realized, such as the full-adder. In Fig. \ref{fig:cadder}, an XOR gate of $A$ and $B$ is created by connecting two Ressler-Feynman switch gates in parallel. Using VCMA-based synchronization, this XOR output is then synchronized with $C_{in}$ before passing through two additional switch gates and a synchronizer to result in $Sum$ and $C_{out}$. 
\FloatBarrier
\section{Physical Reversibility of Skyrmion Logic}

The inclusion of convergent skyrmion paths to perform the OR operation in both the the skyrmion AND/OR gate and full-adder preclude these gates from being physically reversible.  One indication of this is that if a gate is run in reverse, its outputs do not always return to their input positions; this implies information -- and therefore energy -- loss.  The next step toward total physical reversibility is to implement logic gates which may be reversed, with information carriers returning from primary \textit{outputs} to primary \textit{inputs}.

\begin{figure*}[]
    \centering
    \includegraphics[width=\textwidth]{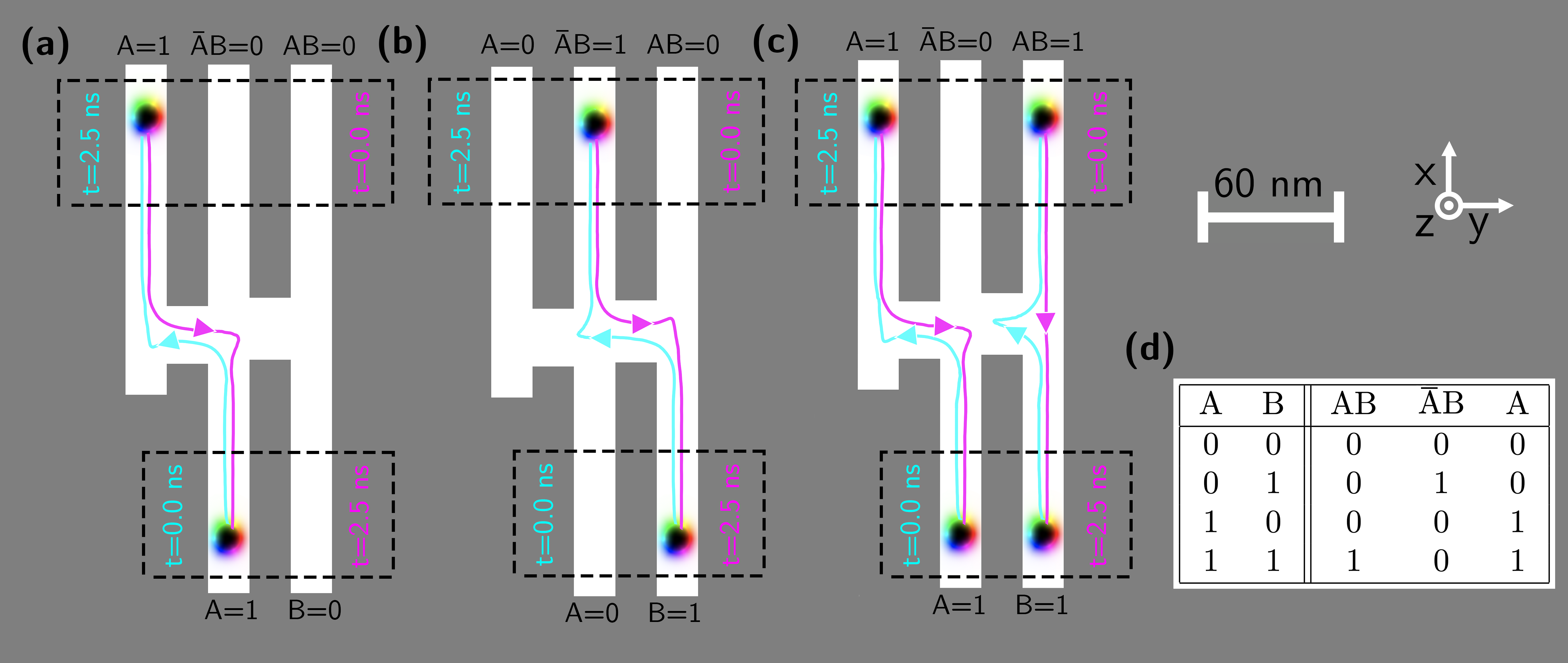}
    \caption{Ressler-Feynman switch gate with current ($\pm5\times10^{10}A/m^2$). Logic shown for input combinations (a) A=1, B=0; (b) A=0, B=1, (c) A=1, B=1. Skyrmion trajectory under positive and negative current is represented by cyan and magenta colored paths, respectively. (d) Ressler-Feynman switch gate truth table}
    \label{fig:Switch}
\end{figure*}

As seen in Fig. \ref{fig:Switch}, the skyrmion Ressler-Feynman switch gate achieves this step towards reversibility: when there is one skyrmion in either input track (Fig. \ref{fig:Switch}c,d), the skyrmion deviates in the $+x$ direction into the adjacent leftward track due to the skyrmion-Hall effect. When both input skyrmions are present (Fig. \ref{fig:Switch}e), the skyrmions repulse each other, forcing the B skyrmion to remain in the right track. Likewise, under reverse current conditions, the $-x$ deviation causes the skyrmions to return to their respective inputs, demonstrating reversibility in this (limited) sense.

As the Ressler-Feynman switch gate is functionally complete, switch gates can be cascaded to implement any arbitrary Boolean function, including a physically reversible full-adder.  Therefore, large-scale skyrmion logic circuits that can be reversed may be constructed from these Ressler-Feynman switch gates, enabling a path toward total physical reversibility.


\section{Conclusions}

Although the proposed scheme can exhibit logical reversibility, more work is needed to achieve total physical reversibility. In particular, at present: (a) the forward and reverse trajectories differ slightly, exhibiting hysteresis, which is a hallmark of dissipative processes; (b) the explicit synchronization operations discard physical information in the form of accumulated timing uncertainty; and (c) the driving currents will necessarily be dissipative unless superconducting materials can be used. However, research on this technology is still in its infancy, and it holds potential for further improvement.





\section*{Acknowledgement}

This work is supported in part by the Advanced Simulation and Computing (ASC) program at the U.S. Department of Energy’s National Nuclear Security Administration (NNSA). Sandia National Laboratories is a multi-mission laboratory managed and operated by National Technology and Engineering Solutions of Sandia, LLC, a wholly owned subsidiary of Honeywell International, Inc., for NNSA under contract DE-NA0003525. This document describes objective technical results and analysis. Any subjective views or opinions that might be expressed in this document do not necessarily represent the views of the U.S. Department of Energy or the United States Government.  Approved for public release, SAND2022-3205 O.

This research is supported in part by the National Science Foundation under CCF award 1910800 and the Texas Analog Center of Excellence undergraduate internship program. 


\end{document}